\documentclass[aps,twocolumn,epsf,floats,pre,superscriptaddress,nofootinbib]{revtex4-1}

\usepackage{graphicx}
\usepackage{amsmath}
\usepackage{microtype}
\usepackage{tikz}
\usepackage{enumitem}
\usepackage{amsfonts}
\usepackage{hyperref}
\usepackage{mathtools}
\usepackage{scrtime}
\usepackage[dont-mess-around]{fnpct}
\usepackage{import}
\usepackage{float}
\usepackage{etoolbox}
\usepackage{placeins}

\newcommand{\beqa}{\begin{eqnarray}}
\newcommand{\eeqa}{\end{eqnarray}}

\newcommand{\xxi}{\boldsymbol{\xi}}

\newcommand{\beq}{\begin{equation}}
\newcommand{\eeq}{\end{equation}}

\def\bk{\boldsymbol{k}}
\def\br{\boldsymbol{r}}
\def\bv{\boldsymbol{v}}

\begin{document}

\title{Discrete Laplacian thermostat for flocks and  swarms: \\ the fully conserved Inertial Spin Model}

\author{Andrea Cavagna}
\affiliation{Istituto Sistemi Complessi (ISC-CNR), Via dei Taurini 19, 00185, Rome, Italy}
\affiliation{Dipartimento di Fisica, Sapienza Universit\`a di Roma, P.le Aldo Moro 2, 00185, Rome, Italy}
\affiliation{INFN, Unit\`a di Roma 1, 00185 Rome, Italy}
    
\author{Javier Crist\'i­n} 
\email{javier.cristin@uab.cat}
\affiliation{Istituto Sistemi Complessi (ISC-CNR), Via dei Taurini 19, 00185, Rome, Italy}
\affiliation{Dipartimento di Fisica, Sapienza Universit\`a di Roma, P.le Aldo Moro 2, 00185, Rome, Italy}
 
\author{Irene Giardina}
\affiliation{Dipartimento di Fisica, Sapienza Universit\`a di Roma, P.le Aldo Moro 2, 00185, Rome, Italy}
\affiliation{Istituto Sistemi Complessi (ISC-CNR), Via dei Taurini 19, 00185, Rome, Italy}
    \affiliation{INFN, Unit\`a di Roma 1, 00185 Rome, Italy}
    
\author{Tom\'as S. Grigera}
\affiliation{Instituto de F\'\i{}sica de L\'\i{}quidos y Sistemas Biol\'ogicos CONICET -  Universidad Nacional de La Plata,  La Plata, Argentina}
\affiliation{CCT CONICET La Plata, Consejo Nacional de Investigaciones Cient\'\i{}ficas y T\'ecnicas, Argentina}
\affiliation{Departamento de F\'\i{}sica, Facultad de Ciencias Exactas, Universidad Nacional de La Plata, Argentina}
\affiliation{Istituto Sistemi Complessi (ISC-CNR), Via dei Taurini 19, 00185, Rome, Italy}
    
\author{Mario Veca}
\affiliation{Dipartimento di Fisica, Sapienza Universit\`a di Roma, P.le Aldo Moro 2, 00185, Rome, Italy}


\begin{abstract}
Experiments on bird flocks and midge swarms reveal that these natural systems are well described by an active theory in which conservation laws play a crucial role.
By building a symplectic structure that couples the particles' velocities to the generator of their internal rotations (spin),
the Inertial Spin Model (ISM) reinstates a second-order temporal dynamics that captures many phenomenological traits of flocks and swarms.
The reversible structure of the ISM predicts that the total spin is a constant of motion, the central conservation law 
responsible for all the novel dynamical features of the model.
However, fluctuations and dissipation introduced in the original model to make it relax, violate the spin conservation law, so that the ISM aligns with the biophysical phenomenology only within finite-size regimes, beyond which the overdamped dynamics characteristic of the Vicsek model takes over.
Here, we introduce a novel version of the ISM, in which the irreversible terms needed to relax the dynamics strictly respect the conservation of the spin. We perform a numerical investigation of the fully conservative model, exploring both the fixed-network case, which belongs to the equilibrium class of Model G, and the active case, characterized by self-propulsion of the agents and an out-of-equilibrium reshuffling of the underlying interaction network. 
Our simulations not only capture the correct spin wave phenomenology of the ordered phase, but they also yield dynamical critical exponents in the near-ordering phase that agree very well with the theoretical predictions.
\end{abstract}

\maketitle

\section{Introduction} \label{sec_intro}

Approximately three decades ago, Vicsek and coworkers enlisted a sizeable number of unwilling  statistical physicists into the fast-growing field of biophysics by introducing a simple little model that elegantly captured the key traits of collective animal  behaviour \cite{vicsek1995novel}. This model, partly rooted in the physics of ferromagnets and critical phenomena, partly drawing inspiration from previous efforts in the field of computer graphics \cite{reynolds1987flocks}, demonstrated that local alignment could induce collective motion and long-range correlations even in active systems; in fact, the Vicsek model showed that activity favours order, allowing for a coherent (or flocking) phase also in two dimensions \cite{tonertu1995}, where equilibrium physics would prevent long-range ordering.  Under more than one respect (but not all respects), this was the start of active matter physics \cite{ramaswamy2010mechanics, ginelli2016, chate2020}.
Vicsek's model is truly remarkable for its simplicity and great generality, so much so that it can be considered the Ising model of active matter. However, as soon as experimental data on actual biological groups started to flow in, it was clear that the quantitative description of certain biological features required models to incorporate some additional elements. Among the many variants and evolutions of the Vicsek model aimed at better describing biological reality \cite{tu1998, gregoire2004, chate2008collective, chate2008a, moreno2020, chate2020, chepizhko2021revisiting, gonzalez-albaladejo2024}, one in particular is of interest for us here: the so-called Inertial Spin Model (ISM) \cite{attanasi2014information, cavagna2015flocking}. 

Observations on starling flocks revealed that collective information during turns propagates linearly and almost without damping throughout the group \cite{attanasi2014information}. Not much later, experiments on natural swarms of midges showed that the dynamical correlation functions display unmistakable traits of inertial relaxation \cite{cavagna2017swarm}; at the same time, swarms have a dynamical critical exponent lower than any other reference ferromagnetic model, and even than the Vicsek model itself \cite{cavagna2017swarm,cavagna2023natural}. All these traits prompted the development a new model in which the inertial features of the dynamics ---  washed out in the Vicsek model --- were reinstated, giving rise to a conservation law able to account for several new biological pieces of phenomenology. This model is the ISM, developed in \cite{attanasi2014information, cavagna2015flocking} and further studied in  \cite{ha2019, markou2021, benedetto2020, ko2023, huh2022}, the overdamped limit of which is --- notably --- the Vicsek model.

The ISM successfully reproduces several traits of natural flocks and swarms, including: the linear propagation law of orientational waves in bird flocks, even in density-homogeneous systems \cite{attanasi2014information}; the quantitative dependence of the speed of propagation of these waves on polarization \cite{attanasi2014information};  the precise inertial shape of the dynamical correlation functions in midge swarms \cite{cavagna2017swarm}; dynamic scaling laws  \cite{cavagna2017swarm}; and finally the precise prediction of the dynamical critical exponent in natural swarms \cite{cavagna2023natural}, \textcolor{black}{to our knowledge the first case of agreement between theory, experiments and numerics on a critical exponent of an actual biological system.} But the ISM has an  issue: despite the fact that its founding pillar is the spin conservation law, the ISM has in fact no way to {\it exactly} respect that very conservation law, which is somewhat annoying.

The reason why that happens is simple. The conservation law of the ISM stems from its novel reversible dynamics, which reinstates the symplectic structure between generalized coordinate and conjugated momentum. However, in order to turn this structure into a {\it relaxational} statistical model, one needs to introduce fluctuation-dissipation terms (noise and friction), which --- in their current form --- \textcolor{black}{cause spin dissipation, thus violating the conservation law. The situation is thus the following:  the experimentally determined dynamic critical exponent is that of the active version \cite{cavagna2023natural} of Halperin and Hohenberg's model G \cite{HH1977}.  The ISM, strictly speaking, is not in this universality class, as spin dissipation drives it, in the thermodynamic limit, to an active but overdamped renormalization group fixed point corresponding to Toner and Tu's field theory \cite{tonertu1998}, i.e.\ the universality class of the Vicsek model.  However, the symplectic structure of the reversible part of the ISM dynamics is important: the field theory that describes the ISM at the mesoscopic scale is that of active model G with the addition of spin dissipation \cite{cavagna2023natural}.  The active model G fixed point is still a valid fixed point in the dissipative theory, but spin dissipation introduces an unstable direction that drives it to overdamped dynamics in the thermodynamic limit.  The Renormalization Group (RG) flow is thus such that there is a crossover with the nontrivial consequence that for finite systems with small friction, the active model G fixed point rules the behaviour (yielding the experimental dynamic critical exponent).  The crossover to overdamped dynamics is only observed for very large systems.}

\textcolor{black}{Thus, up to now, the ISM was able to reproduce underdamped inertial dynamics only in finite-size systems and for small friction.  This is quite acceptable in practice, and in line with the experimental observations, since the overdamped behaviour is only reached for sizes beyond reasonable numbers for biological groups.  But from a purely theoretical point of view this is unpleasant; we would like to have a microscopic model in the universality class of self-propelled Model G: able to relax (thermalize) and \emph{at the same time} respect the conservation law responsible for the new physics, irrespective of system size.  Here we introduce a variation of the ISM in which total spin is exactly conserved at all times, irrespective of the value of the kinetic coefficient and of the temperature, and thus irrespective of the phase (ordered or disordered).  Thanks to this, the new model is the first one proposed to date that exactly belongs to the universality class of self-propelled Model G at all length scales.}

\section{The fully-conserved Inertial Spin Model}
\label{sec:cons}

\subsection{Deterministic structure}

The novelty of the model we are going to introduce will be in the new kind of irreversible and yet conservative terms. However, in order to better explain the evolution from one model to another, we will start our discussion from the --- rather unphysical --- purely deterministic cases and gradually build up noise and dissipation.

For all the self-propelled models described in this work the first equation of motion is
$\dot{\boldsymbol r}_i = \boldsymbol v_i$, hence we will not repeat it every time; if/when we will consider the fixed-network version of a model, we will simply set $\dot{\boldsymbol  r}_i = 0$. The equation for the velocity, on the other hand, defines the particular model; we will start from the simplest one, that is Vicsek.

\subsubsection{The Vicsek rule}

In what follows we will consider fixed modulus of the velocity, and to simplify the notation we will set $|\boldsymbol v_i|=1$ (the reader should be careful then in comparing with the previous literature, where the fixed value of the speed is often left as a dimensional parameter, $v_0$).
In its barest, deterministic form, Vicsek dynamics \cite{vicsek1995novel} in continuous time (more useful for our developments), can be written as,
\begin{equation}
\eta \frac{d\boldsymbol v_i}{dt}  = \left(\boldsymbol v_i \times J \sum_j n_{ij}\boldsymbol  v_j \right) \times \boldsymbol v_i \ ,  
\label{eq:vicsek-det}
\end{equation}
where  $J$ corresponds to the alignment strength and $n_{ij}$ is the adjacency matrix ($n_{ij}=1$ if the agents $i$ and $j$ interact, $n_{ij}=0$ otherwise).  The interacting pairs can be chosen with a metric or topological criterion (which one to use may depend on the system under consideration \cite{cavagna2018correlations}).  Of course, in the active case where the equation $\dot{\boldsymbol r}_i = \boldsymbol v_i$ is at work, the adjacency matrix depends on time, which is at the heart of the out-of-equilibrium nature of models of self-propelled particles.  \textcolor{black}{Note that Eq.~\eqref{eq:vicsek-det} applies to the continuous-time version of the Vicsek model with periodic boundary conditions.  Some differences in behaviour have been reported, for the discrete time Vicsek model, depending on whether periodic, reflecting or open (confining field) boundary conditions are employed \cite{chepizhko2021revisiting,gonzalez2023scale,chate2020}.}

The double cross product is a compact way to express the conservation of the modulus of $\boldsymbol v_i$.  The neighbours of agent $i$ exert a so-called {\it social force}, $\boldsymbol f_i= J \sum_j n_{ij} \boldsymbol v_j$; the term $(\boldsymbol v_i \times \boldsymbol f_i) \times \boldsymbol v_i$ is the projection $\boldsymbol f_i^\perp$ of $\boldsymbol f_i$ onto the plane orthogonal to $\boldsymbol v_i$.  Thus the effect of the Vicsek rule is to \emph{rotate} particle $i$'s velocity towards the average direction of motion of its neighbours.  The way this happens is {\it direct}, i.e.\ the force $\boldsymbol f_i$ itself rotates the velocity $\boldsymbol v_i$ through the double cross product.\footnote{Note that the rule $\dot{\boldsymbol v}_i =\boldsymbol v_i \times \boldsymbol f_i$ also conserves $|\boldsymbol v_i|$.  However, it does not represent an alignment interaction, but a {\it precession} of $\boldsymbol v_i$ around $\boldsymbol f_i$, rather than a rotation of $\boldsymbol v_i$ {\it towards} $\boldsymbol f_i$.}
This dynamic rule creates an overdamped dynamics for the velocity, which is exactly what the ISM wants to change.  Note that we have left the dissipation parameter $\eta$ in the equation (which at this level simply defines a time scale).  Although usually eliminated by rescaling time in the Vicsek model, it will be convenient to keep it for a comparison with the ISM in what follows.

\subsubsection{The Inertial Spin Model rule}

The key idea of the ISM --- compared to the Vicsek model --- is that the social alignment force, instead of directly rotating the velocity of the agent, acts on its {\it spin}, which is the generator of the rotations of the velocity.  In turn the spin rotates the velocity, according to \cite{cavagna2015flocking}
\begin{subequations}
  \begin{align}
    \frac{d\boldsymbol v_i}{dt} & = \frac{1}{\chi}\boldsymbol s_i \times \boldsymbol v_{i}, \\
    \frac{d\boldsymbol s_{i}}{dt} & =  \boldsymbol v_i \times J \sum_j n_{ij} \boldsymbol v_j  \ .
  \end{align}
  \label{eq:ISM-det}
\end{subequations}
This symplectic relationship between velocity and spin is the same as that between position and angular momentum, when rotations in the external space are considered; from this analogy we clearly see that the parameter $\chi$ acts as a generalized inertia, i.e.\ a social or physiological resistance of the agents to rotate their direction of motion.  \textcolor{black}{It can be shown that, as long as the interaction matrix is symmetric, the equation of motion for the velocity can also be derived from a (time-dependent) Lagrangian \cite{benedetto2020}.}

The crucial point is that, if the interaction is {\it symmetric} under a given transformation, a global conservation law emerges  due to Noether's teorem. This is the case of the alignment interaction: the social force $\boldsymbol f_i$ is invariant under rotation of the velocity, so that --- once the symplectic structure of velocity and spin is restored --- the generator of rotations, namely the total spin of the system is, {\it in absence of dissipative forces}, a constant of motion.  This conservation law is a game-changer: in the ordered phase of the ISM it is the cause of the emergence of linear spin waves observed in actual biological flocks \cite{attanasi2014information}, while in the near-ordering phase it accounts for much of the dynamical phenomenology of natural swarms of insects, including inertial dynamic relaxation \cite{cavagna2017swarm} and an anomalous dynamical critical exponent \cite{cavagna2023natural}.

\subsection{Stochastic structure}

\subsubsection{Adding fluctuation and dissipation}

These models must be complemented with some stochastic rule accounting for the fact that biological agents are unlikely to obey the interaction rule in a perfect deterministic way, and (most importantly) allowing the relaxation of the dynamics.  \textcolor{black}{Relaxation is achieved by coupling the system to a thermal bath; this coupling must necessarily include two terms: a fluctuating (stochastic noise) force which injects energy into the system, and a dissipative (friction) term, which transfers energy to the bath.  We refer to these terms as fluctuation-dissipation terms, or irreversible terms.}
In the Vicsek case, it is enough to add to the r.h.s.\ of \eqref{eq:vicsek-det} a noise term (taking care of preserving the constraint $|\boldsymbol v_i|=1$), because dissipation is already included in the overdamped nature of the model. One way to do this is
\begin{equation}
\eta \frac{d\boldsymbol v_i}{dt}  = \left(\boldsymbol v_i \times J \sum_j n_{ij} \boldsymbol v_j \right)\times\boldsymbol v_i
 + \boldsymbol \xi_i \times \boldsymbol v_i \ .  
\label{eq:vicsek}
\end{equation}
where $\boldsymbol \xi_i$ is a white noise with correlator,
\begin{equation}
\label{eq:noise-corr}
\langle \xi_i^{\mu}(t) \xi_j^{\nu}(t') \rangle = 2 \eta T \ \delta(t-t')\delta_{ij}\delta^{\mu\nu} \ .
\end{equation}
The choice of noise in \eqref{eq:vicsek} is not the only possibility; one could also add a stochastic term of the form $\boldsymbol v_i \times( \boldsymbol \xi_i \times \boldsymbol v_i) =  \boldsymbol \xi_i^\perp$, similarly to the projection of the force.  This is how the continuous-time Vicsek model is written in \cite{cavagna2015flocking}.  In fact, both types of noise conserve the modulus of the velocity and have the same correlator (given by the orthogonal projector), hence from a practical point of view both are equivalent, but the form in \eqref{eq:vicsek} is more convenient because it will appear naturally when we will take the overdamped limit of the inertial model.

In ISM, we need to be more careful: in which equation should we add the noise? Because noise is a stochastic (generalized) {\it force,} and forces act on the time derivative of the conjugate momentum to the coordinate (in this case, the velocity is an internal coordinate), it is reasonable to add noise in the second equation of~\eqref{eq:ISM-det}.  Moreover, it is natural to associate to each fluctuating force, an irreversible force causing relaxation (dissipation) in the  same equation.  If we follow this scheme we obtain the ISM equations (compare to \eqref{eq:ISM-det}),
\begin{subequations}
  \begin{align}
    \frac{d \boldsymbol v_i}{dt} & = \frac{1}{\chi}\boldsymbol s_i \times \boldsymbol v_{i}, \\
    \frac{d\boldsymbol s_{i}}{dt} & =  \boldsymbol v_i \times J \sum_j n_{ij} \boldsymbol v_j - \frac{\eta}{\chi} \boldsymbol s_i + \boldsymbol \xi_i,
  \end{align}
  \label{eq:ISM}
\end{subequations}
where the noise obeys \eqref{eq:noise-corr}.  Since the noise enters in the equation for the spin, there is no need to project it to keep the norm of $\boldsymbol v_i$ fixed  \textcolor{black}{(this equation is simpler than the corresponding equation first introduced in \cite{cavagna2015flocking}; the difference has no strong consequences and this form is more convenient for the present purposes, see App.~\ref{sec:cond-vdots}).}
The dissipation and noise are the irreversible terms that allow the system to relax towards a non-equilibrium stationary state.  The nice thing about this structure is that the overdamped limit of this model (very large $\eta$) corresponds to the Vicsek model (see App.~\ref{sec:overdamped-limit-ism}).

\subsubsection{The need for fully conservative fluctuation-dissipation}

\textcolor{black}{The fact that the overdamped limit of the ISM is the Vicsek model acquires particular relevance in the light of the RG, as it turns out that the parameter $\eta$ \emph{always grows} under an RG transformation (unless it is exactly zero).  This means that at very large length scales the system will be described by the Vicsek model (actually its continuous version, namely Toner and Tu's field theory) even if the microscopic dynamics is that of ISM (provided $\eta>0$).  How, then, is it that the ISM is at all useful for the description of actual biological groups?  Cutting a rather long \cite{cavagna2019short, cavagna2019long, cavagna2020equilibrium, cavagna2023natural} story short, the answer is finite size: although the statement about the overdamped limit ruling the large length scales is true, if the microscopic $\eta$ is small to begin with, the overdamped limit is reached only for very large scales --- larger than the system size.  In this case, the observed behaviour is that of the inertial dynamics, i.e.\ ISM.  The RG explains this in terms of the existence of two fixed points: one (unstable) at $\eta=0$, the other (stable) at $\eta=\infty$.  For small $\eta$, the RG flow stays near the $\eta=0$ fixed point, and only after many iterations of the RG transformation it finally goes to $\eta=\infty$.  Since each iteration of the RG implies looking into the system from a larger scale, for small systems this last crossover never takes place, and the unstable fixed point with $\eta=0$ rules the dynamics.}

\textcolor{black}{So, for theoretical investigation we would like a microscopic model that is described by the $\eta=0$ fixed point at all scales (i.e.\ that belongs to its universality class).  Setting $\eta=0$ in~\eqref{eq:ISM} means that there is neither dissipation nor noise, i.e.\ the model reduces to the {\it deterministic} version~\eqref{eq:ISM-det}: the system cannot relax, and we lose temperature as a parameter that can e.g.\ tune the amount of order.  To make statistical-mechanical sense of the model with $\eta=0$ we need fluctuation-dissipation terms that are conservative, making the theory stochastic and relaxational even when $\eta=0$ \cite{cavagna2019short, cavagna2019long}.}

\textcolor{black}{Interestingly, the RG itself generates under coarse-graining the terms we need (seeApp.~\ref{sec:issue-dissipation-rg}).  At the mesoscopic (field theory) level, these are actually well-known, as they arise e.g.\ in field theories with conservative dynamics such as model G for antiferrmagnets \cite{HH1977}. In short, looking at the coarse-grained spin $\boldsymbol s(x,t)$, instead of a dissipative equation like
\begin{equation}
\frac{\partial \boldsymbol s(x,t)}{\partial t} = \mathrm{reversible \ terms} -\eta \boldsymbol s(x,t) + \boldsymbol \xi \ ,
\label{baccalu}
\end{equation}
with noise correlator
\begin{equation}
\langle \xi_\alpha(x,t) \xi_\beta(x',t') \rangle = 2\eta \ \delta(x-x') \delta(t-t') \delta_{\alpha\beta}  ,
\label{pippazzu}
\end{equation}
one writes
\begin{equation}
\frac{\partial \boldsymbol s(x,t)}{\partial t} = \mathrm{reversible \ terms} + \lambda\nabla^2 \boldsymbol s(x,t) + \boldsymbol \xi \ ,
\label{baccala}
\end{equation}
with
\begin{equation}
\langle \xi_\alpha(x,t) \xi_\beta(x',t') \rangle = - 2\lambda\nabla^2 \ \delta(x-x') \delta(t-t') \delta_{\alpha\beta} .
\label{pippazza}
\end{equation}
In Fourier space $-\lambda\nabla^2 \to \lambda k^2$, and it is therefore clear that the space integral of the spin field is automatically conserved by the new irreversible $\lambda$-terms.  In the next section we thus seek equivalent fluctuation-dissipation terms for the discrete model.}

\subsection{The fully conservative Inertial Spin Model}

\textcolor{black}{We now present the new variant of the ISM, which is \emph{irreversible} (because it is coupled to a thermal bath) but exactly conserves the total spin. In this sense the model is \emph{conservative irreversible,} and we refer to the new coupling to this spin-conserving thermal bath as fully conservative fluctuation-dissipation terms.}

\subsubsection{The Discrete Laplacian}

To write the discrete equivalent of \eqref{baccala}-\eqref{pippazza} we can use the discrete Laplacian operator, which is defined as \cite{gross2018graph}, 
\begin{equation}
\Lambda_{ij} = -n_{ij}+\delta_{ij}\sum_k n_{ik} \ , 
\end{equation}
where $n_{ij}$ is the adjacency matrix (notice that the discrete Laplacian is a positive-definite matrix, i.e. $\Lambda \sim -\nabla^2$). The general zero mode of the Laplacian is present of course also in its discrete version,
\begin{equation}
\sum_i\Lambda_{ij}=0 \ .
\end{equation}
In perfect similarity to \eqref{baccala}-\eqref{pippazza} we can then write the conserved relaxational dynamics for the discrete spins as,
\begin{equation}
\frac{d\boldsymbol s_{i}}{dt} =  \boldsymbol v_i \times  J\sum_j n_{ij} \boldsymbol v_j - \frac{\lambda}{\chi} \sum_j \Lambda_{ij} \boldsymbol s_{j} + \boldsymbol \xi_i \ .
\label{glip}
\end{equation}
with,
\begin{equation}
\langle \xi_i^{\mu}(t) \xi_j^{\nu}(t') \rangle = 2 \lambda \Lambda_{ij}T \delta(t-t')\delta_{\mu\nu}.
\label{banzai}
\end{equation}
From \eqref{glip}-\eqref{banzai} we have,
\begin{equation}
\frac{d\boldsymbol{S}}{dt} = \sum_i\frac{d \boldsymbol{s}_i}{dt} = \sum_i \boldsymbol{\xi}_i \ ,
\end{equation}
but from \eqref{banzai} we see that the random variable $\sum_i \boldsymbol{\xi}_i$ has zero mean and zero variance, hence it must be identically zero for each realization of $\boldsymbol{\xi}_i$, finally proving that these new fluctuation-dissipation terms conserve the total spin, $\dot S=0$.

\begin{figure}[t]
\centering
\includegraphics[width=0.40\textwidth]{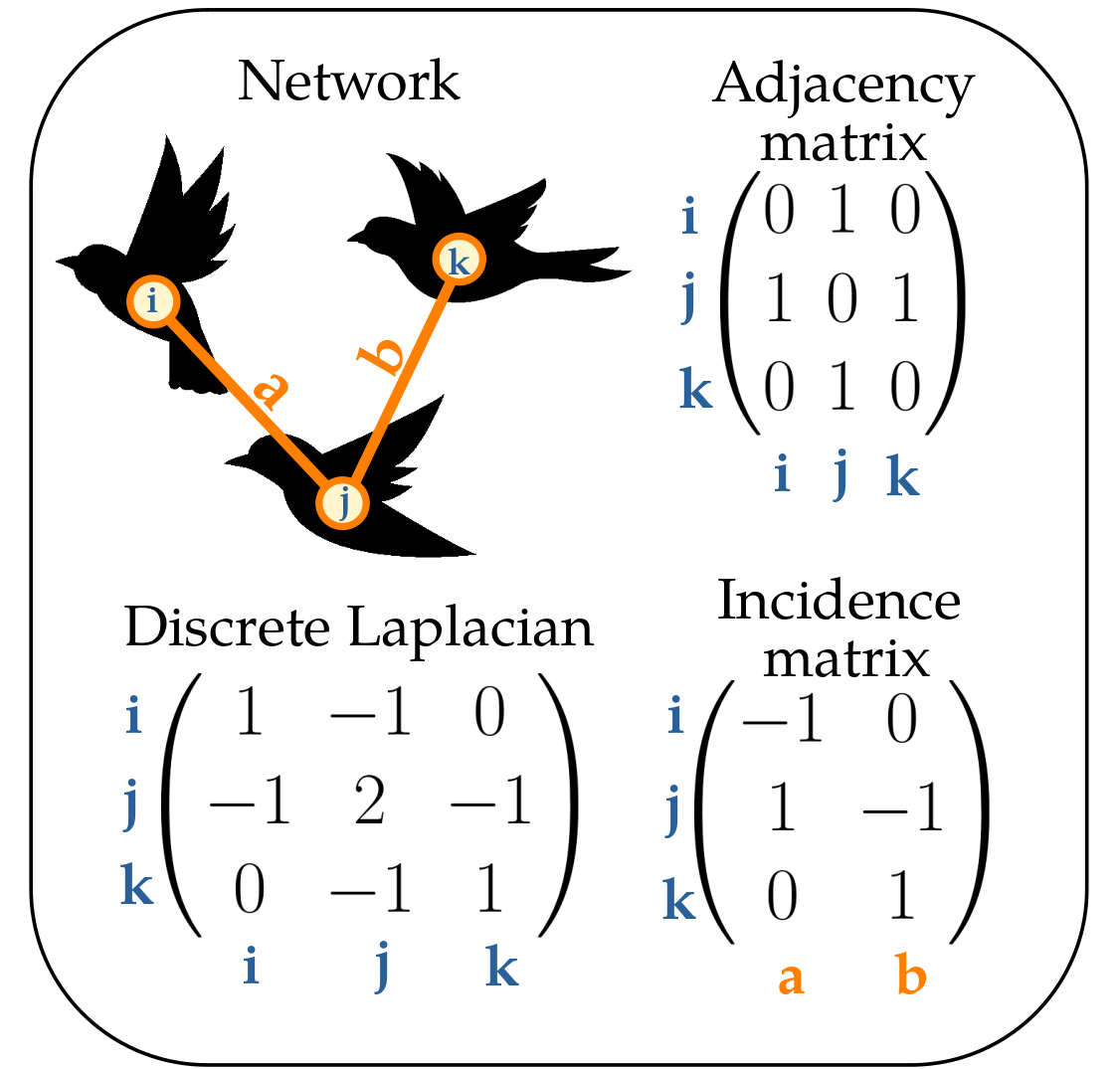}
\caption{Schematic view of the adjacency matrix $n$, incidence matrix $D$, and discrete Laplacian $\Lambda$ in a very simple network.}
\label{fig:scheme}
\end{figure}

\subsubsection{From noise on the sites to noise on the links}
\label{nana}
To produce a noise $\xxi_i$ whose correlator is proportional to $\Lambda_{ij}$, we can use the discrete Laplacian thermostat introduced in \cite{cavagna2023discrete} and further developed in \cite{cavagna2023noise}: we must pass from a noise defined on the sites to a noise defined on the links.
Let us label the sites of the lattice with $\lbrace i,j,\dots\rbrace$ and the links with $\lbrace a, b,\dots\rbrace$. The {\it incidence matrix}, $D_{ia}$, is constructed as follows \cite{gross2018graph}: after arbitrarily assigning a direction to each link $a$, we set $D_{ia}=+1$ if $i$ is at the end of $a$,  $D_{ia}=-1$ if $i$ is at the origin of $a$, and $D_{ia}=0$ if site $i$ does not belong to $a$ (Fig.\ref{fig:scheme}); note that, by construction, we have, 
\beq
\sum_{\mathrm{sites\ } i} D_{ia} = 0 \ .
\label{piscotto}
\eeq
Notice also that the arbitrariness in the definition of $D_{ia}$ due to the arbitrary choice of the directions of the links, reflects the inevitable arbitrariness in the definition of the derivative on a general discrete lattice.\footnote{Notice that, in the active case, the arbitrary assignment of the $\pm1$ sign to the links can be done once and for all at the beginning of the simulation, irrespective of the changing mutual positions of the particles.}
The crucial property of the incidence matrix is that its square {\it over the links} is equal to the discrete Laplacian \cite{gross2018graph}, 
\begin{equation}
\sum_{\mathrm{links\ } a} D_{ia}D^\mathrm{T}_{aj} = \Lambda_{ij}  \ .
\label{zimbra}
\end{equation}
If we define on each link $a$ a $\delta$-correlated Gaussian noise, $\boldsymbol{\epsilon}_a$, with variance, 
\begin{equation}
\langle {\epsilon}^\mu_a(t) {\epsilon}^\nu_b(t') \rangle = 2T \, \lambda\; \delta_{ab} \,\delta_{\mu\nu} \, \delta(t-t') \ ,
\label{pino}
\end{equation}
we can now build the site noise acting on each spin $i$ as,
\begin{equation}
\boldsymbol{\xi}_i(t)= \sum_a D_{ia} \, \boldsymbol{\epsilon}_a(t) \ ,
\label{zumba}
\end{equation}
whose variance is,
\begin{eqnarray}
 \langle{\xi}^\mu_i(t) {\xi}^\nu_j(t') \rangle &=&
\sum_{ab} D_{ia} D_{jb} \, \langle \epsilon^\mu_a(t) \epsilon^\nu_b(t') \rangle
\nonumber
\\
 &=& \sum_{a} D_{ia} D^\mathrm{T}_{aj} \, 2T \;  \lambda \,\delta_{\mu\nu} \, \delta(t-t') 
\nonumber
 \\
&=&2T \;  \lambda\,  \Lambda_{ij} \,\delta_{\mu\nu} \, \delta(t-t') \ ,
\end{eqnarray}
so that we recover exactly the desired expression, equation \eqref{banzai}. Because by construction $\sum_i D_{ia} = 0$, from \eqref{zumba} we have that,
\beq 
\sum_i \boldsymbol{\xi}_i = 0 \ , 
\eeq
which makes it even more apparent the fact that the new noise conserves the total spin in \eqref{glip}.


\subsubsection{The new model}
\label{sec:new-model}

To summarize, the new fully conservative ISM model is defined by the equations,
\begin{equation}
\begin{split}
\frac{d\boldsymbol r_i}{dt} & = \boldsymbol v_i \\
\frac{d\boldsymbol v_i}{dt} & = \frac{1}{\chi}\boldsymbol s_i \times \boldsymbol v_{i} \\
\frac{d\boldsymbol s_{i}}{dt} & = \boldsymbol  v_i \times J\sum_j n_{ij} \boldsymbol v_j  - \frac{\lambda}{\chi} \sum_j \Lambda_{ij}\boldsymbol  s_{j} + \boldsymbol \xi_i \ , 
\end{split}
\label{eq:cISM_ac}
\end{equation}
where the noise is built from equations \eqref{pino} and \eqref{zumba}, and therefore satisfies,
\begin{equation}
\langle \xi_i^{\mu}(t) \xi_j^{\nu}(t') \rangle = 2 \lambda \Lambda_{ij}T \delta(t-t')\delta_{\mu\nu}.
\end{equation}
To distinguish this modified version from the standard ISM, we refer to it as the Fully Conserved Inertial Spin Model (FC-ISM). As described inApp. \ref{sec:cond-vdots}, the new Laplacian terms violate the condition $\boldsymbol s_i \cdot \boldsymbol v_{i}=0$, which we therefore decided to relax. The stochastic dynamics anyway keeps it true on average.

We notice that the fact of having introduced a conservative fluctuation-dissipation part, does not prevent us from adding {\it also} a dissipative part to the dynamics, 
\begin{subequations}
\begin{align}
\frac{d\boldsymbol r_i}{dt} & = \boldsymbol v_i , \\
\frac{d\boldsymbol v_i}{dt} & = \frac{1}{\chi}\boldsymbol s_i \times \boldsymbol v_{i} ,\\
\frac{d\boldsymbol s_{i}}{dt} & =  \boldsymbol v_i \times J\sum_j n_{ij} \boldsymbol v_j - \frac{1}{\chi}\sum_j(\eta\, \delta_{ij} + \lambda\,  \Lambda_{ij}) \boldsymbol s_{j} + \boldsymbol \xi_i  ,
\end{align}
\label{eq:cISM_ac+eta}
\end{subequations}
with the corresponding noise,
\begin{equation}
\langle \xi_i^{\mu}(t) \xi_j^{\nu}(t') \rangle = 2 (\eta\, \delta_{ij} + \lambda\, \Lambda_{ij}) \,T \delta(t-t')\delta_{\mu\nu}.
\end{equation}
We will not study such general model here, but it is important to understand that in principle and in practice, one could; we cannot exclude that a certain amount of {\it bona fide} dissipation of the spin is present in natural systems, due to the interaction between the agents and the environment, so the possibility to have both $\eta$ and $\lambda$ in the model gives it more realism and flexibility.  Also, in the model with both $\eta$ and $\lambda$ we can give a dimensional argument to understand the existence of the crossover length ${\cal R}_c$ (see~\ref{sec:crossover-length}).

\section{Results}

To delve into the exploration of the novel fully conservative model, we conduct numerical analyses of equations \eqref{eq:cISM_ac} in both the Fixed Network (FN) and Self-Propelled Particles (SPP) scenarios. We begin by examining the FN case. We simulate the model on a fixed lattice, i.e.\ without self-propulsion: particles sit on a lattice and the equation $d\boldsymbol r_i/dt = \boldsymbol{v}_i$ is ignored, so that $\boldsymbol{v}_i$ is just an orientational degree of freedom not related to the velocity.  In this case the model is in the universality class of Halperin and Hohenberg's Model G \cite{HH1977}, a class that also includes the quantum antiferromagnet.  This is a good starting point to test the model, since Model G is well-known, and we know what to expect:  we know the exact values of the critical exponents near criticality, and we anticipate the linear propagation of information in the form of spin waves in the low-temperature phase.  Once the static equilibrium case has been characterized, we will reintroduce the self-propulsion in the dynamics,  $\dot{\boldsymbol{r}}_i=\boldsymbol{v}_i$, which causes the underlying interaction network to undergo constant reshuffling.  In this scenario, the model falls in the universality class of self-propelled Model G \cite{cavagna2023natural}, for which we have clear RG predictions in the near-ordering phase.  While we lack a theoretical prediction for the low-temperature case, observations from starling flocks and from previous numerical simulations suggest that we will continue to observe a linear propagation of information.

\subsection{Numerical details}

We have employed a 4th-order Runge-Kutta (RK) method to integrate the deterministic part of equations \eqref{eq:cISM_ac}.  \textcolor{black}{The stochastic part was integrated with an Euler scheme, as in ref.~\cite{cavagna2023discrete}.   This method ensures that the cross product strictly preserves the norm of $\bv_i$, and avoids numerical errors in the deterministic part that could interfere with spin conservation for small $\lambda$.}  We set $|\bv_i|=1$ and time step $\Delta t=5 \times 10^{-4}$, which ensures that all simulations realized for this work converge.

For the FN case, the simulations have been conducted on a cubic lattice of size $L$ with periodic boundary conditions (PBC) and unit inter-particle distance, so that the total number of particles is $N=L^3$.  Each site has $n_c=2d=6$ neighbours.  We have simulated system sizes from $L=4$ to $L=18$, with a number of samples depending of $L$. 

For the SPP case, the simulations have been conducted on a cubic box of size $L$ (also with PBC). The total number of particles is $N$; $N$ and $L$ have been chosen to have a density $\rho=\frac{N}{L^3}=1$ for all cases. We have simulated system sizes that span from from $L=4$ to $L=14$, with a number of samples depending of $L$.  The particles interact with a metric rule: $i$ and $j$ interact if they are separated by less than a given distance $r_c$. In terms of the adjacency matrix $n_{ij}$, this reads
\begin{align}
\begin{cases}
     n_{ij}=1, &  \text{if } r_{ij} \leq r_c,  \\
    n_{ij}=0, &  \text{if } r_{ij} > r_c  .
    \end{cases}
    \label{eq:metric}
\end{align}

We aim to simulate systems with homogeneous density. We choose $r_c=1.6$  as in \cite{cavagna2023natural}, resulting in an average of $17$ particles in the interaction neighborhood. A higher number of interacting neighbors helps reduce density fluctuations. \textcolor{black}{Another approach to maintain density homogeneity is normalizing the social force   in Eq.~\eqref{eq:ISM} by local density \cite{chepizhko2021revisiting}. This is achieved by adjusting $Jn_{ij} \to J\frac{n_{ij}}{n_i}$, with $n_{i}=\sum_j n_{ij}$. This procedure, used in \cite{cavagna2023natural}, leads to non-reciprocal forces and violates total spin conservation. Thus, we normalize the social force as $J \to J/{\frac{1}{2}(n_{i}+n_{j})}$, ensuring reciprocity and conserving total spin.}

 

\textcolor{black}{For both FN and SPP we have set $\chi=1$.  The values of $J$, $\lambda$ and $T$ are specified for each result.} We have carefully tested that all samples reach a stable steady state (i.e. thermalize in the FN case).  The initial velocities \textcolor{black}{were chosen in random directions, except in the  SPP-ordered case, were all velocities were initialized to the same random direction in order to reach the steady state more quickly.}  The spins have been initialized as $\boldsymbol{s}_i=0, \forall i$.  In the SPP case, the initial positions $\br_i$ have been chosen to be the sites of a cubic lattice.

\begin{figure}[t]
\centering
\includegraphics[width=0.5\textwidth]{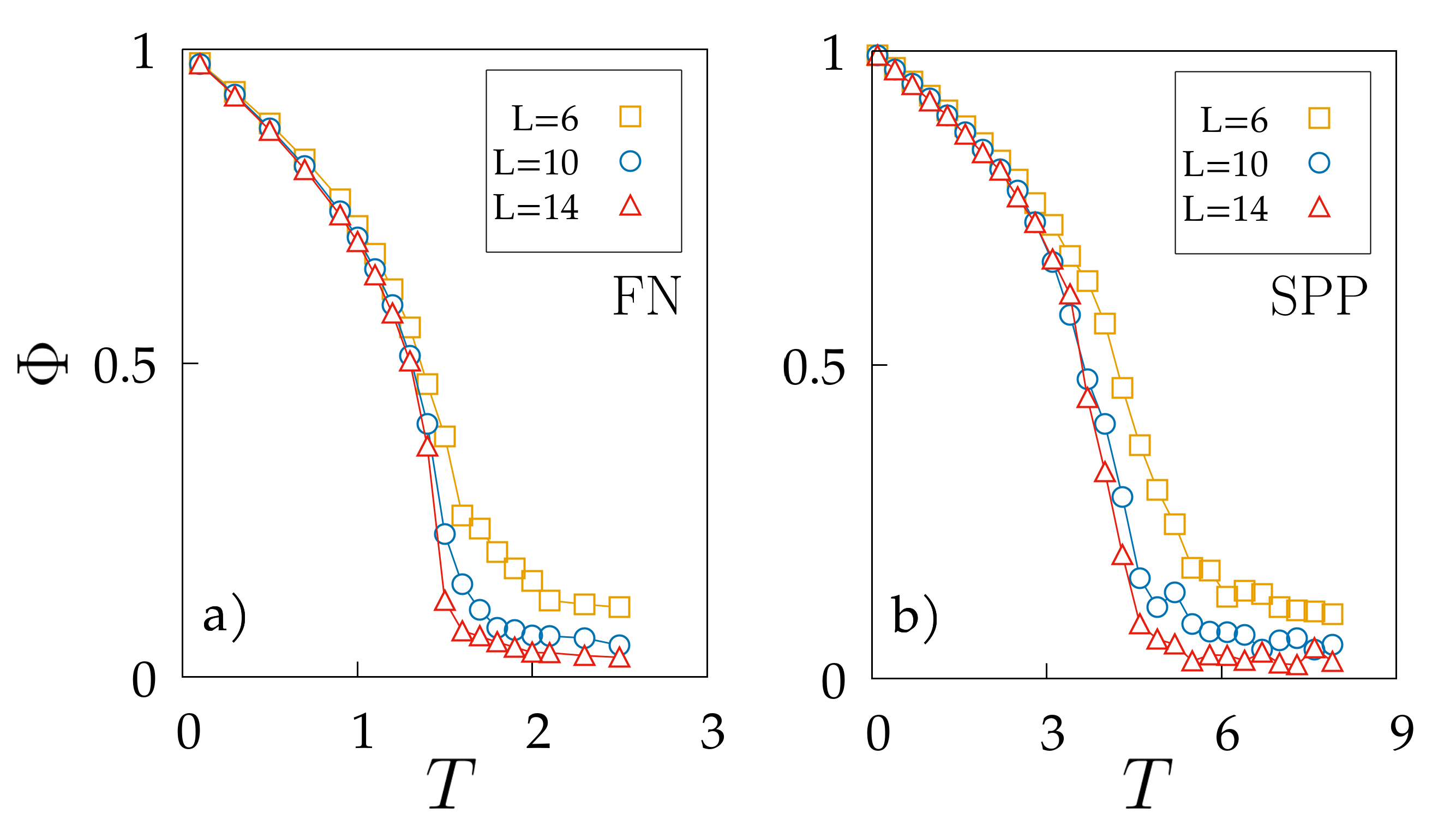}
\caption{Stationary order parameter $\Phi$ when the temperature $T$ is tuned for different system sizes for a) the FN case (with $\lambda=0.5$) and b) the SPP case (with $\lambda=0.1$).}
\label{fig:pd}
\end{figure}

We have computed the order parameter
\begin{equation}
  \label{eq:OP} 
  \Phi = \frac{1}{N}  \left\lvert \sum_i  \frac{\boldsymbol{v}_i}{\lvert \boldsymbol{v}_i\rvert} \right\rvert,
\end{equation}
that allows us to characterize the static phase diagram (Figure \ref{fig:pd}), and thus to identify the temperature regime where the system is in the ordered and disordered phases, separated by the critical temperature $T_c$.

We have also  computed the dynamic connected correlation function of the velocities,
\begin{equation}
  \label{eq:correlation-fourier}
    C(\bk,t) =   \frac{1}{N}  \left\langle \sum_{i,j}^N \delta\bv_i(t_0)\cdot\delta\bv_j(t_0+t) e^{-i\bk\cdot \br_{ij}}\right\rangle_{t_0} , 
\end{equation}
where $\delta \bv_i(t) = \bv_i(t)-(1/N)\sum_k\bv_k$  , $\br_{ij}=\br_i(t_0) - \br_j(t_0+t)$ and $\langle\cdot\rangle_{t_0}$ represents an average over the time origin $t_0$.  We have \textcolor{black}{averaged the correlation functions of the three orthogonal directions of the wave-vector $\boldsymbol{k}$,}
\begin{equation}
  \label{eq:correlation-fourier-iso}
    C(k,t) =\frac{1}{3}\left(  C(\bk_x,t) +C(\bk_y,t)+C(\bk_z,t) \right),
\end{equation}
where $\bk_x=(k,0,0)$, $\bk_y=(0,k,0)$ and $\bk_z=(0,0,k)$.  \textcolor{black}{In the FN case where the correlation function is isotropic, this serves to improve statistics.  In the SPP case this is an approximation, since isotropy does not hold; see discussion below.} We have finally computed the correlation function $C(k,\omega)$, defined as the temporal Fourier transform of the correlation function $C(k,t)$.

\subsection{Fixed-Network test of the new model}

\subsubsection{FN --- Near-ordering phase}
\label{FN_Tc}

Classic RG studies of the dynamic critical exponent of Model G  give $z=1.5$ \cite{HH1977}, a value in full agreement with experiments on perovskites as RbMnF$_3$ \cite{tucciarone1971quantitative} and with numerical simulations of the Heisenberg antiferromagnet \cite{landau1999spin, nandi2019nonuniversal,cavagna2023discrete}. Here we verify that the fully conserved ISM in the Fixed Network case falls in the same universality class.
Calculating the critical exponent $z$ requires using dynamic scaling,
\textcolor{black}{
\begin{equation}
  C(k,t) = C_0(k) f(t/\tau_k;k\xi ),
  \label{bubu0}
\end{equation}
where $C_0(k)$ is the static correlation function and the relaxation time at wavevector $k$ is given by}
\begin{equation}
    \tau_k=k^{-z}g(k\xi),
\label{bubu}
\end{equation}
with $g$ a scaling function and all the dependence on temperature $T$ is contained goes in the correlation length $\xi$.

We can obtain the characteristic correlation time $\tau_k$ from the dynamic correlation functions, either in the time domain or in the frequency domain. Following the standard definition in \cite{HH1969scaling}, the characteristic frequency $\omega_k$ can be calculated from the following relation, 
    \begin{equation}
    \int_{-\omega_{k}}^{\omega_{k}} \frac{d \omega} {2 \pi}\frac{C(k,\omega)}{C_0(k)}=\frac{1}{2}.
    \label{eq:HH_wc}
\end{equation}
The characteristic frequency $\omega_k$ is naturally the inverse of the relaxation time $\tau_k$, which --- translating  \eqref{eq:HH_wc} to the time domain --- is given by
\begin{equation}
   \int_0^\infty \!\! \frac{dt}{t} \, \frac{C(k,t)}{C(k,t=0)}
   \sin\left(\frac{t}{\tau_k}\right) = \frac{\pi}{4}.
   \label{eq:HH_t}
\end{equation}
In order to exploit Eq.~\eqref{bubu} and to compute the dynamic exponent, it is necesary to identify first the critical region.  From the static correlation function $C(k)=C(k,t=0)$ we obtain the susceptibility $\chi = C(k_\text{max})$, where $k_\text{max}$ is the wave-vector of the first maximum of $C(k)$\footnote{
In the usual case, where the static connected correlation is defined using phase-averages, the susceptibility is proportional to $C(k=0)$. Here, however, we compute the connected correlation by subtracting the \emph{space} average of the velocity. With this definition $C(k=0)$ vanishes by construction, due to the sum rule $\sum_i\delta\bv_i=0$;  on the contrary, $C(k_\text{max})$ is well defined and it can be shown to provide a good finite size proxy for the susceptibility --- see \cite{cavagna2018correlations} for a discussion of this point.}.
For any given size $L$, we can then identify the finite-size critical temperature $T_c(L)$, as the temperature where the susceptibility  is maximal.  This temperature defines the critical region, where the correlation length is proportional to the system size, i.e. $\xi(T_c(L)) \sim L$. Working in this regime and choosing $k=1/\xi \sim 1/L$, Eqs.~\eqref{bubu0} and \eqref{bubu} give
\begin{equation}
   C(k,t) = C_0(k) f(t/\tau;1), \qquad    \tau \sim L^z .
\label{bibi}
\end{equation}

\begin{figure}[t!]
\centering
\includegraphics[width=0.5\textwidth]{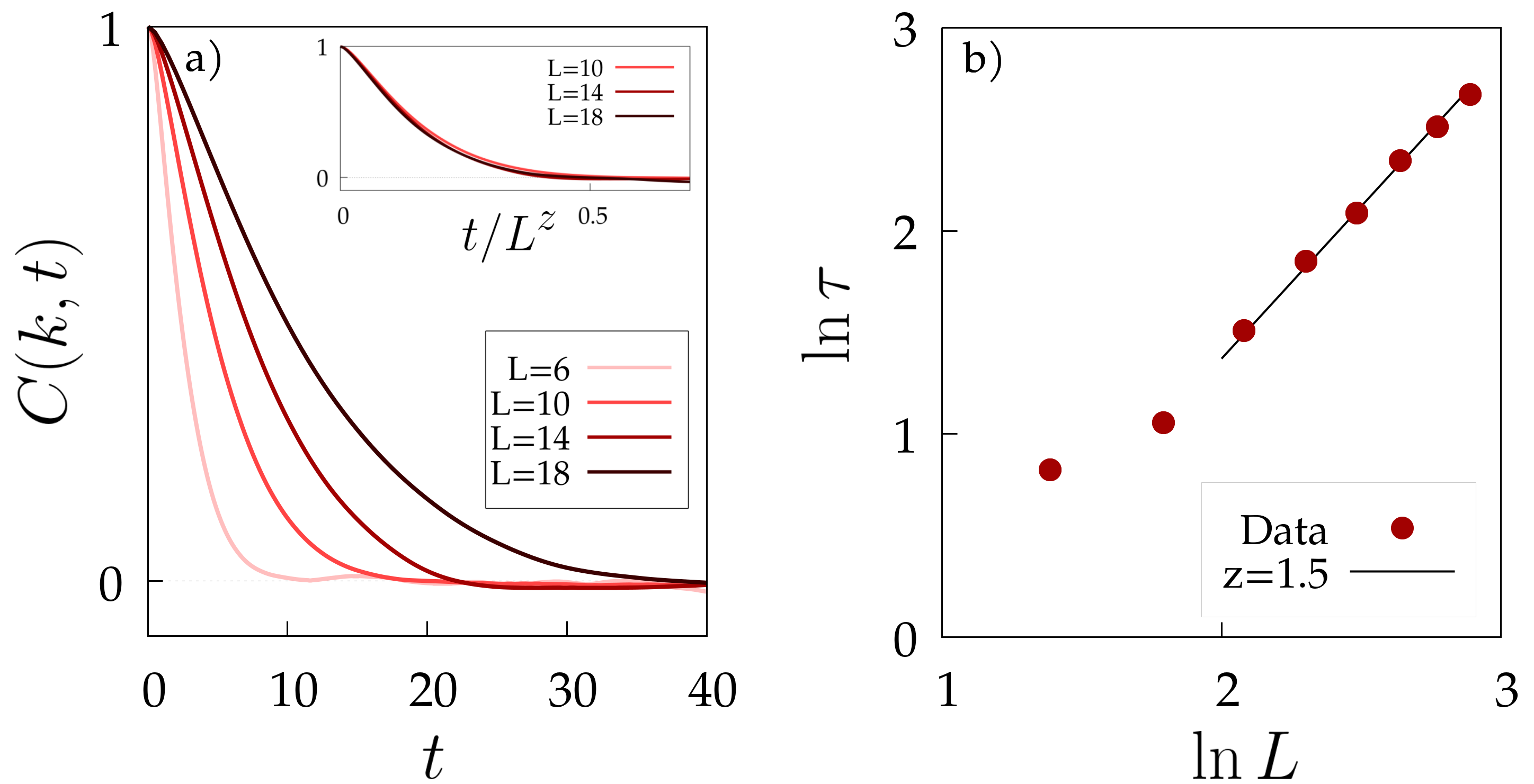}
\caption{{\bf Fixed-network --- near ordering:} a) Correlation functions $C(k,t)$ as a function of time, at the pseudo-critical temperature $T_c(L)$ for different sizes $L$, \textcolor{black}{$J=1$} and $\lambda=0.5$. The correlation functions are normalised by their value at $t=0$, and they are evaluated at $k=2\pi /L$ (see main text).  \textcolor{black}{Inset: scaling plot of the normalised time correlation functions with scaling variable $t L^{-z}$, which shows the validity of scaling form \eqref{bibi}.} b) Relaxation time $\tau$ as a function of the system’s size $L$.  The line indicates the fit with the exact dynamical critical exponent, $z = 1.5$, over the six largest sizes. }
\label{fig:FN_Tc}
\end{figure}

Eq.~\eqref{bibi}  provides a straightforward method to compare our simulations with the theoretical prediction of Model G, where $z=1.5$.  In Figure \ref{fig:FN_Tc}a) we display several $C(k,t)$ curves, normalized by their value at $t=0$, obtained  at the critical temperature for different system sizes. For each one of these curves, we then compute the global relaxation time following Eq.~\eqref{eq:HH_t}, and plot it as a function of $L$. A fit of these data gives us the numerical estimate of the dynamical critical exponent $z$.  The result is displayed in panel b) of Figure~\ref{fig:FN_Tc},  and it matches quite well the $z=1.5$ value.  \textcolor{black}{The inset of panel a) shows the normalised correlation function plotted against the scaling variable $t L^{-z}$, confirming the validity of the scaling Eq.~\eqref{bibi}.}  For thoroughness, we also performed a fitting procedure with a free exponent $z_{\text{fit}}$ using the six largest sizes, yielding $z_{\text{fit}} = 1.48 \pm 0.04$. Our results therefore confirm that the universality class of the fully conserved ISM in the fixed network (i.e. equilibrium) case is the same as that of the microscopic quantum antiferromagnet, namely the universality class of Model G.

\subsubsection{FN --- Ordered phase}
\label{FN_Tlow}

At low temperatures ($T \ll T_c$), the system exhibits spontaneous symmetry breaking. In this regime, Model G exhibits linear propagation of information in the form of spin waves. This characteristic behavior has been verified for the microscopic quantum antiferromagnet \cite{landau1999spin,cavagna2023discrete}. In the next paragraphs, we analyze and discuss information propagation and the spin-wave  behaviour within the context of the fixed-network FC-ISM.

We note that, due to symmetry breaking, all velocities $\bv_i$ predominantly point in the same mean direction $\boldsymbol{V}=(1/N) \sum_i \bv_i$. It is then convenient to rewrite the individual velocities in terms of fluctuations: $\bv_i=\boldsymbol{V} + \delta \bv_i^L + \boldsymbol{\pi}_i$, where $\boldsymbol{\pi}_i \cdot \boldsymbol{V}=0$ and - at low temperature - $|\boldsymbol{\pi}_i| \ll |\boldsymbol{V}|$ and  $|\delta \bv_i^L| \sim O(|\boldsymbol{\pi}_i|^2)$. By rewriting the equations of motion in terms of the variables $\{\boldsymbol{\pi}_i\}$ and assuming that $\boldsymbol{V}$ is constant in time, we can expand up to first order, obtaining a linear dynamics for  $\boldsymbol{\pi}_i$. This procedure is known as the spin-wave expansion. The resulting equation reads 

\begin{equation}
    \frac{d^2 \boldsymbol{\pi}_i}{dt^2}= - \frac{J}{\chi} \sum_j \Lambda_{ij} \boldsymbol{\pi}_j - \frac{\lambda}{\chi}\sum_j \Lambda_{ij} \frac{d\boldsymbol{\pi}_j}{dt} + \boldsymbol{\hat n} \times \boldsymbol{\xi}_i ,
\end{equation}
where $\boldsymbol{\hat n}$ is the collective direction of motion. In the continuum limit, when $\Lambda_{ij}\to - \nabla^2$, the propagator (Green's function) of the homogeneous differential equation can be easily computed in Fourier space giving the following dispersion relation for its poles,
\begin{equation}
\Tilde{\omega}(k)=-\frac{\lambda k^2}{2 \chi}i \pm \sqrt{\frac{J k^2}{\chi}-\left(\frac{\lambda k^2}{2 \chi}\right)^2}.
\label{eq:disp}
\end{equation}

The imaginary part of $\Tilde{\omega}$ encodes the decay in time of the $C(k,t)$.
The real part of $\Tilde{\omega}$, that we denote here as $\omega_c$, represents the characteristic frequency of oscillation of the correlation function $C(k,t)$ (or the spin waves frequency). The characteristic frequency $\omega_c(k)$ is,
\begin{equation}
\omega_c(k)= \sqrt{\frac{J k^2}{\chi}-\left(\frac{\lambda k^2}{2 \chi}\right)^2}.
\label{eq:wc}
\end{equation}
\begin{figure}[t!]
\centering
\includegraphics[width=0.5\textwidth]{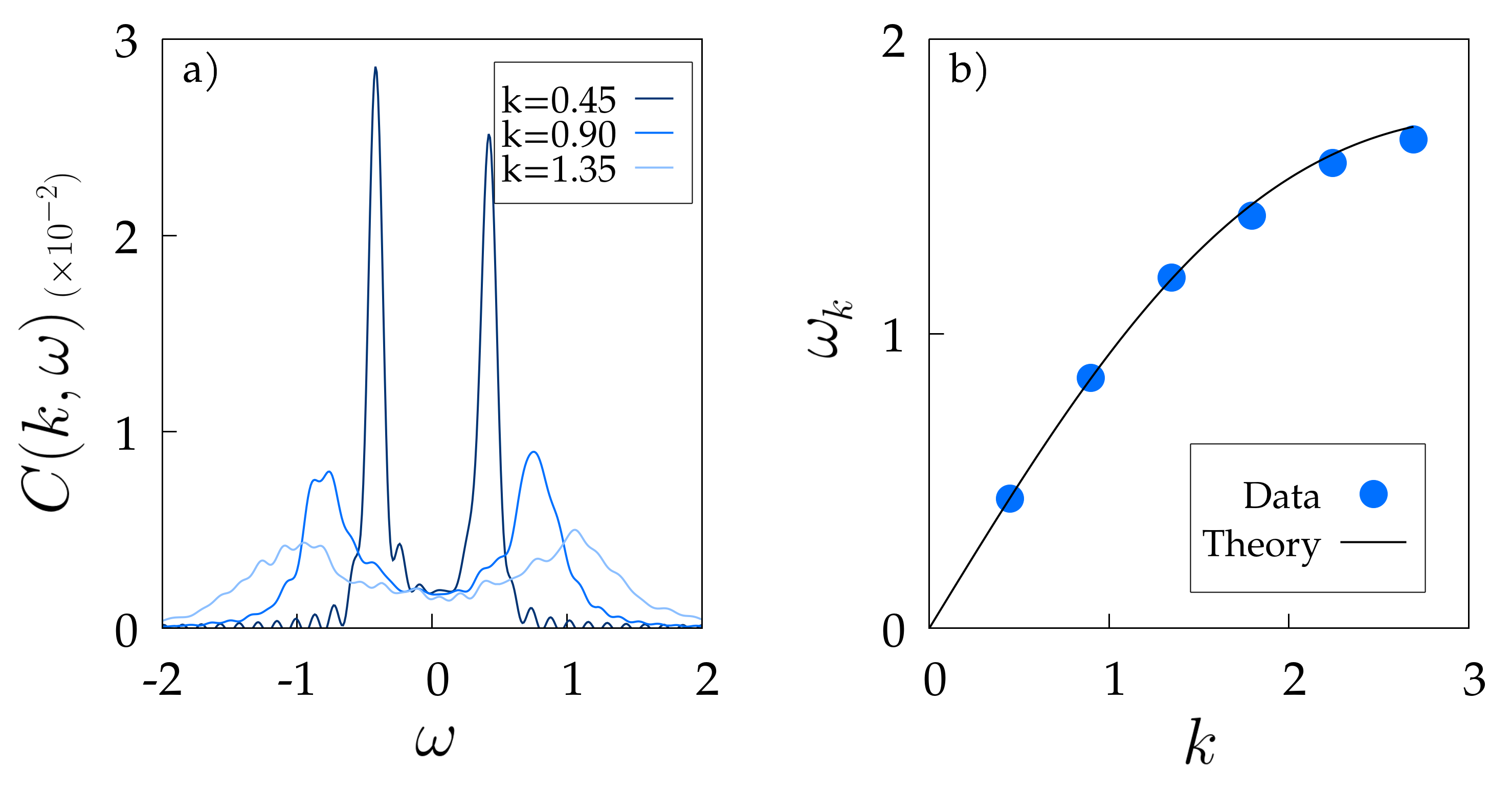}
\caption{{\bf Fixed network --- ordered phase:} a) Correlation functions $C(k,\omega)$ for different wave-vectors $k$ for a system of size $L=14$ with  $T = 0.2$, \textcolor{black}{$J=1$} and $\lambda=0.5$.  b) Characteristic frequency $\omega_k$ computed from Eq.~\eqref{eq:HH_wc} for different values of $k$ (blue points). The continuous curve corresponds to the characteristic frequency of the spin-waves $\omega_c(k)$ derived analytically, see Eq.~\eqref{barzotta}.
}
\label{fig:FN_lowT}
\end{figure}

In fact, however, we work on a discrete cubic lattice with periodic boundary conditions (PBC), for which the eigenvalues of the Laplacian operator are not $k^2$ but rather $4 \sin^2(k/2)$; hence, the actual dispersion relation for our simulation should be, 
\begin{equation}
    \Tilde{\omega}(k)=-\frac{\lambda 4 (\sin{\frac{k}{2}})^2}{2 \chi}i \pm \sqrt{\frac{J   4(\sin{\frac{k}{2}})^2}{\chi}-\left(\frac{\lambda 4 (\sin{\frac{k}{2}})^2}{2 \chi}\right)^2} ,
    \label{eq:disp_disc}
\end{equation}
and hence,
\begin{equation}
\omega_c (k)= \sqrt{\frac{J   4(\sin{\frac{k}{2}})^2}{\chi}-\left(\frac{\lambda 4 (\sin{\frac{k}{2}})^2}{2 \chi}\right)^2}
\label{barzotta}
\end{equation}
The most direct way to analyze the characteristic frequency of oscillation in our simulations is by focusing on $C(k,\omega)$. 
The correlation functions $C(k,\omega)$ obtained from our simulations distinctly reveal the presence of a smooth peak at a specific frequency (see Figure \ref{fig:FN_lowT} a)). Notably, the position of this peak varies with the value of $k$. 
To compute numerically the characteristic frequency for different values of $k$, we resort to Eq.~\eqref{eq:HH_wc}.
Indeed it can be easily shown that, in presence of a dominant mode, $\omega_k$ tends to the frequency of that mode \cite{HH1969scaling}.  In Figure \ref{fig:FN_lowT} b), we then compare the characteristic frequencies obtained directly from the simulations with the analytical expression $\omega_c(k)$ derived from the spin-wave expansion \eqref{barzotta}. In conclusion, the numerical data confirm the presence of propagating spin waves in the fixed-network fully conserved ISM and they are in full agreement with the theoretical predictions of the spin-wave expansion.

\begin{figure}[t!]
\centering
\includegraphics[width=0.5\textwidth]{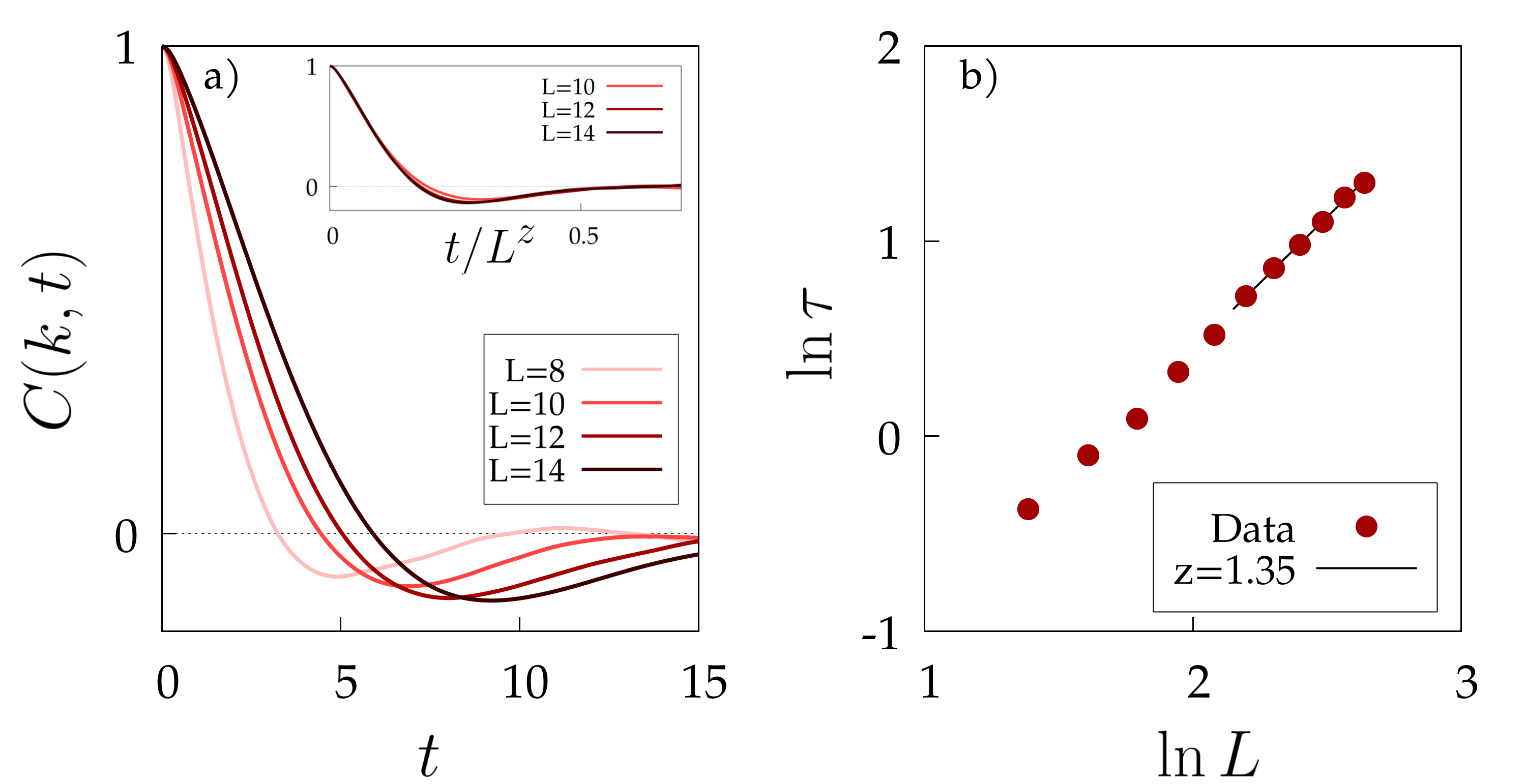}
\caption{{\bf Self-propelled case --- near ordering:} a) Correlation functions $C(k,t)$ at the pseudo-critical temperature $T_c(L)$ for different sizes $L$, \textcolor{black}{$J=18$ (with normalised interactions, see text) and  $\lambda=0.1$}.  The correlation functions are normalized by their value in $t=0$, and they are evaluated at $k=2\pi /L$.  \textcolor{black}{Inset: scaling plot of the time correlation with scaling variable $t L^{-z}$.}  b) Relaxation time $\tau$ as a function of the system’s size $L$. The line indicates the fit with the theoretical dynamical critical exponent, $z = 1.35$, over the six largest sizes.}
\label{fig:SPP_Tc}
\end{figure}

\subsection{Self-propelled test of the new model}

\subsubsection{SPP --- Near-ordering phase}
\label{SPP_Tc}

The dynamic critical exponent of the self-propelled Model G is known to be $z=1.35$, which is compatible with the estimate of $z$ from experimental data of midge swarms \cite{cavagna2023natural}.  As also discussed in \cite{cavagna2023natural}, numerical simulations of the standard ISM model (i.e., $\lambda=0$), performed with sufficiently small $\eta$, also yield an exponent $z$ consistent with $1.35$. However, as discussed previously, the standard ISM is quite limited if we want to explore the conservative universality class. Indeed, it exhibits this class behaviour only for 
$L \ll {\cal R}_c(\eta)$ so that for larger sizes it requires increasingly smaller values of $\eta$, which in turn requires increasingly longer simulation times.  Here, on the other hand, we will determine $z$ in simulations of the purely conservative ISM, where these problems do not arise.

To obtain $z$, we follow the same procedure used for the FN case (Sec.~\ref{FN_Tc}).  In Figure \ref{fig:SPP_Tc}a)  we display the correlation functions $C(k,t)$ at the critical temperature, for several values of $k$. At short times, we can observe the effect of inertia (i.e. a decay that is not exponential), much in the same way as for the fixed network case. After the initial decay, however, there appears to be an oscillation even at the near-ordering temperature, which was not present on the fixed lattice. We currently lack a clear explanation for this phenomenon; this is a non-equilibrium model for which we do not have a full theoretical prediction other than the RG determination of $z$. However, we can put forward an educated guess. In systems with a low-temperature spin wave phase (antiferromagnets and, more generally,  the Model G class), it is well known that some precursors of the low-$T$ phase emerge already around the critical temperature \cite{marshall1965critical, doniach1966, berk1966}.  Such precursors (called ``paramagnons'' in the modern condensed matter literature) show themselves as unexpected real parts of the characteristic frequency, in a phase where one would only expect an imaginary part; hence, in the time domain, they are simply extra oscillations in the $C(k,t)$.  The specific form of paramagnons is known to be highly dependent on the specific details of the material/model under consideration, hence it may very well be that the out-of-equilibrium version of the fully conserved ISM merely has a more pronounced abundance of paramagnons compared to its equilibrium counterpart. A more careful investigation of the numerical data and of the background theory are required to clarify this point; this is  left for future research.

Whatever the specific shape of $C(k,t)$, we can calculate the largest relaxation time $\tau$ from  $C(k,t)$ at the pseudo-critical temperature $T_c(L)$ and plot the obtained relaxation times $\tau$ vs. $L$.  This is exactly what we show in panel b) of Fig.~\ref{fig:SPP_Tc}: our results show that our simulations are in complete agreement with the prediction of the self-propelled Model G, i.e, $z=1.35$.  For the sake of completeness, we  fit  a free exponent to the six largest sizes, obtaining  an exponent $z_{fit}=1.35 \pm 0.04$.  \textcolor{black}{The scaling plot (inset of panel a) shows a very good collapse and again confirms the validity of dynamic scaling.}  We remark that this is the first determination of the exponent $z$ from simulations falling in the universality class of self-propelled Model G without the need to exploit a finite-size crossover, nor (which is the same) the need to tune any specific parameter (i.e. the friction $\eta$, as it occurs in the standard ISM). 
\begin{figure}[t!]
\centering
\includegraphics[width=0.5\textwidth]{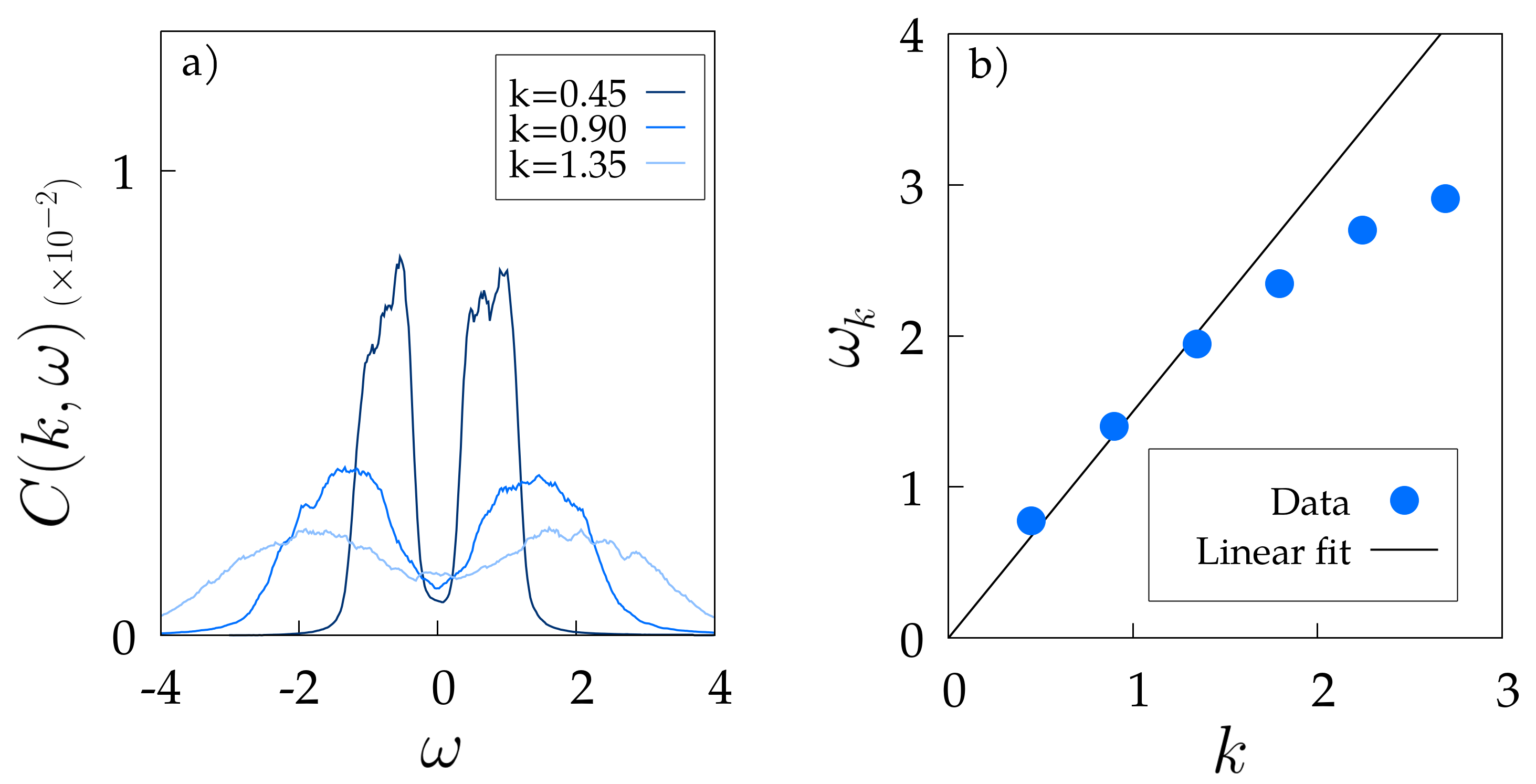}
\caption{{\bf Self-propelled case --- ordered phase} a) Correlation functions $C(k,\omega)$ for different wave-vectors $k$ for a system of size $L=14$ with $T = 0.5$, \textcolor{black}{$J=18$ (with normalised interactions, see text) and $\lambda=0.1$}. b) Characteristic frequency $\omega_k$ computed from Eq.~\eqref{eq:HH_wc} for different values of $k$ (blue points); the straight line is a linear fit of the three lowest-$k$ data points.}
\label{fig:SPP_lowT}
\end{figure}
\subsubsection{SPP --- Ordered phase}
\label{SPP_Tlow}

As in the FN case, the system undergoes spontaneous symmetry breaking in the low temperature regime ($T \ll T_c$). This manifests as a collective alignment of the individual velocities, with most of the particles moving coordinately along the polarization direction. From a biological standpoint, this behaviour mirrors the coordinated motion observed in starling flocks. We notice at this point a difference between our study of the FC-ISM up to now and the real dynamics of starling flocks. Previous studies have highlighted the topological nature of interactions within natural flocks \cite{Ballerini2008}, where individuals interact with their $n_c$ nearest neighbours, rather than with all individuals within a fixed radius $r_c$ (metric interaction), which is what we have implemented thus far into the FC-ISM. The problem is that topological interactions can be non-reciprocal: the fact that one bird interacts with another does not imply that the second bird interacts back with the first ($n_{ij} \neq n_{ji}$). Within the present definition of the fully-conservative model, this fact can potentially violate the conservation law of the spin. Therefore, we opt to retain metric-based interactions here, prioritizing the exploration of conservation law effects over an exact replication of starling flock behaviour. We leave the (perhaps non-trivial) generalization of the fully conservative model to non-reciprocal interactions to future work. 

The procedure we follow to characterize information propagation is the same adopted in Section \ref{FN_Tlow} for the fixed network model. In contrast with the equilibrium case, though, there is no analytical determination, nor any previous theoretical study of the dispersion relation in the active ISM to compare our data with. Yet, we do expect the presence of linearly propagating modes, at least in the low $k$ limit. Indeed, this is what has been observed in experiments on natural starling flocks \cite{attanasi2014information}, which led to development of the ISM in the first place. Besides, numerical simulations of the dissipative ISM, in its low $\eta$ regime, confirm the existence of linearly propagating modes in the system. Moreover, from a theoretical point of view, the fact that the fixed-network counterpart {\it has} linearly propagating modes, together with the observation that in the deeply ordered phase the interaction network $n_{ij}(t)$ changes very slowly in time \cite{mora2016local}, suggests that linear propagating modes of the orientation should survive also in the active regime. Yet, of course, we have no reason to expect that the full dispersion relation in the active case is exactly the same as in the equilibrium one.  \textcolor{black}{ What we do expect, however, is that there exists a regime above a certain value of $\lambda$ in which information does not propagate ($\Tilde \omega$ becomes purely imaginary), as happens in the equilibrium case.  Accordingly we choose $\lambda=0.1$, a very low value we expect to be below this threshold.} 


In Fig.~\ref{fig:SPP_lowT} a), we show the correlation functions $C(k,\omega)$ of the active, fully conservative ISM. 
\textcolor{black}{We notice that the structure of the correlation function is similar to the one of the equilibrium case, Fig.~\ref{fig:FN_lowT}.  There are two peaks, which maximum shifts with the value of $k$. However, the peaks are not as smooth as in the equilibrium case.  There are various possible reasons for this, including lack of statistics and active effects (due to the slow reshuffling of the interaction network).  Another reason may be an underlying anisotropy in  $C(k,\omega)$.  To obtain Fig.~\ref{fig:SPP_lowT} we have averaged over three directions of the wave vector.  This is of no consequence for the equilibrium theory, which is isotropic even in the symmetry-broken phase, but this is not necessarily true for SPPs (for example, anisotropic correlation functions are theoretically expected and observed in the ordered phase of the Vicsek model \cite{tu1998sound}).  If some anisotropic structure were indeed present in the active correlations, then by averaging the three directions we would create
a complicated structure by mixing different propagating modes.  The detailed study of a possible anisotropic structure requires much heavier numerical simulations and it is left to further studies.  In the present analysis we do a simpler characterisation of the ordered phase (more similar to what feasible in an experimental situation where data are scarce), and ask a simpler question, namely
whether the mean characteristic frequency (defined with the half-width method) does have a linear dependence on $k$.  The answer to this question is positive: Fig.~\ref{fig:SPP_lowT}) b) reveals a linear correlation between the frequency $\omega_k$ and $k$ at small momentum, consistently with previous simulations on the dissipative ISM and with the experimental data on starling flocks.  Even though a more detailed study of the low-temperature phase is certainly needed to fully explore the dispersion relation, these results confirm that the spin conservation law is responsible for creating linearly propagating modes of the orientation in the active ordered phase of SPP models.}


\section{Conclusions}

We have presented and tested numerically a version of the inertial spin model with relaxational terms (friction and noise), but where the total spin is nevertheless strictly conserved.  The formalism we have used exploits the recently introduced discrete Laplacian thermostat of \cite{cavagna2023discrete,cavagna2023noise}.  The framework also allows for additional non-conserving spin dissipation (see Eq.~\eqref{eq:cISM_ac+eta}), but we have not tested this more general case in simulations.  The main motivation for the introduction of this model is that it provides a microscopic model belonging to the dynamic universality class of active, inertial and conservative systems (active model G), described by the new RG fixed point recently described in the inertial active field theory of Ref.~\cite{cavagna2023natural}.  The original dissipative ISM, in contrast, is described by this fixed point only at sufficiently low dissipation and sufficiently small system size, being ruled in the thermodynamic limit by an overdamped fixed point.  As we have pointed out, the strict thermodynamic limit is of little use to describe biological groups, which are rather small and can thus be ruled by unstable fixed points in the presence of crossover phenomena.  Nevertheless, it is theoretically desirable to be able to pinpoint at least one microscopic model that is strictly (i.e.\ in the thermodynamic limit) a member of this new universality class.

We have considered first the fixed-lattice (non-active) version of the model.  Since this equilibrium case should belong to the well-known class of Model G, it serves as a test for the model and the numerical procedure.  In the near-critical region we have found the dynamical critical exponent to be fully compatible with $z=1.5$ (the exponent is known to be exactly $z=3/2$ in this case \cite{HH1977}).  In the low-temperature phase, the dynamic correlation function $C(k,\omega)$ shows clear spin-wave peaks and the observed dispersion relation coincides with the theoretical result.

We then simulated the active (self-propelled) model, which is of course out-of-equilibrium.  In this case, the available theoretical results are limited.  In the near-critical regime, we found $z=1.35$, which coincides with the one-loop RG result as well as with the numerical simulations of the weakly dissipative ISM \cite{cavagna2023natural}; incidentally,  this is also the exponent measured in experiments on natural swarms of midges.    Density fluctuations, and their coupling to the order parameter, are the reason why the ordering transition in the Vicsek model is discontinuous in the thermodynamic limit \cite{ramaswamy2010mechanics, toner2012reanalysis, ihle2011, Solon2015, chate2020}.  However, finite-size effects are strong, and  the phenomenology of continuous transitions is observed in not-too-large systems both in simulations \cite{chate2008collective} and in actual biological groups \cite{attanasi2014finite}.  To study this phenomenology theoretically, incompressibility is added as a hard constraint (as e.g.\ in \cite{chen2015critical, cavagna2023natural}).  In numerical studies, the size at which the discontinuous phenomenology starts to be observed depends strongly on the details of the model (such as discrete vs.\ continuous time \cite{chepizhko2021revisiting} or scalar vs.\ vector noise \cite{chate2008collective}).  \textcolor{black}{To study the critical regime of the (dissipative) ISM under conditions of minimal density fluctuations, the simulations of Ref.~\cite{cavagna2023natural} normalized the alignment coupling $J$ by the number of particles within the interaction radius, $J n_{ij} \to J n_{ij}/ n_{i}$, a prescription that is known to make the model less prone to phase separation. Here we have used the symmetric normalisation  $J n_{ij} \to J \frac{n_{ij}}{\frac{1}{2}(n_{i}+n_{j})}$ to preserve spin conservation.}
 
We have also looked for spin waves in the ordered phase of the active model. We have set a low value of $\lambda$ to avoid excessive attenuation of spin waves that could lead to a purely imaginary dispersion relation.   There are no theoretical results to look at for guidance in this case, but the expectation from experimental results in flocks \cite{attanasi2014information}, and simulations on the weakly dissipative ISM \cite{cavagna2015flocking},  is that there should be propagating modes with a linear dispersion relation (second sound).  We have found excitations showing up as peaks in $C(k,\omega)$ that, although somewhat broad, do display a dispersion relation with a linear real part.
However, the phenomenology of the low temperature phase clearly needs further investigation; in particular, the question of a possible anisotropic structure of the correlation functions should be addressed.

Another question that remains open is the following: the ordered phase of ISM is a good candidate to describe flocks of birds; however, it is known that in starling flocks interactions are topological rather than metric \cite{Ballerini2008}.  It is not clear how to deal with topological interactions in this model.  Topological interactions lead to a non-symmetric adjacency matrix (because particle $i$ being $n$-th nearest neighbour of particle $j$ does not necessarily imply the reciprocal relation), thus violating strict spin conservation at the deterministic level.  One could still introduce spin-conserving noise (employing a different matrix for the dissipative part), but it is not clear how this would mix with a non-conserving deterministic interaction, and what difference would it make with respect to the dissipative case.  The topological case calls for a detailed separate study.

Despite the several open questions, this first test of the new conservative model shows that it can be useful for numerical studies of the active, inertial and conservative universality class, with no possible complications due to the presence of the small dissipation, as in the original ISM. An important instance in which the complete absence of dissipation may help in the analysis is the following. As we repeatedly remarked, the complicated structure of spin-wave excitations in the low temperature phase of the active model has yet to be fully understood.  However, by construction, it is clear that this complicated structure cannot be due to the presence of dissipation, as one could have surmised if the original dissipative ISM had been used to simulate the system.  Although this and other issues remain open to further studies, the usefulness of the present model  seems promising.


\section*{Acknowledgements}

This work was supported by ERC Advanced Grant RG.BIO (Contract No. 785932) and by the MIUR (Ministero dell’Istruzione) grants INFO.BIO (Protocol No.
R18JNYYMEY) and PRIN2020 Grant No. 2020PFCXPE. We thank G.~Pisegna and M.~Scandolo for fruitful discussions.

\appendix

\section{On the condition $\boldsymbol s_i \cdot \boldsymbol v_{i}=0$}
\label{sec:cond-vdots}

The equation for the spin in \eqref{eq:ISM} is subtly different from (and simpler than) the corresponding ISM equation first introduced in \cite{cavagna2015flocking}. The reason is the following. In the original model it was imposed the constraint $\boldsymbol s_i \cdot \boldsymbol v_{i}=0$, which required that both the dissipative term and the noise term in the spin dynamics had to be crossed-multiplied by $\boldsymbol v_{i}$. The purpose of the present work, though, is to add a conservative irreversible structure to the model, which can make it stochastic while conserving the total spin.  The new conservative terms (Sec.~\ref{sec:new-model}) violate the condition $\boldsymbol s_i \cdot \boldsymbol v_{i}=0$ (although it still holds on average).  Because of this, we find it more convenient to relax the constraint from the outset, thus introducing this slightly simpler version of the ISM (Eq.~\eqref{eq:ISM}), which is the one we generalise in the Main Text to the conservative case.  From the physical point if view, the noise in equation \eqref{eq:ISM} corresponds to the case in which the system is directly exchanging spin with the environment, while the original ISM equations of \cite{cavagna2015flocking} correspond to the case in which the system exchanges  \emph{the linear generator,} $\boldsymbol w_i $, with $\boldsymbol s_i  = \boldsymbol v_i  \times \boldsymbol w_i $ \cite{carruitero2023inertial} with the environment. Because the ISM is an effective model, with no direct connection with the actual mechanics of the agents described, and because we have no idea of the actual interaction with the environment considered as a heat bath, both descriptions are acceptable, depending on the desired applications of the model. We notice, though, that once we give up the constraint $\boldsymbol s_i \cdot \boldsymbol v_{i}=0$, we can no longer work on the ISM within the framework of the linear generator $\boldsymbol w_i $, as it was recently done in \cite{carruitero2023inertial}.

\section{The overdamped limit of the ISM}
\label{sec:overdamped-limit-ism}

The overdamped limit is always a limit in time; the inertial time scale of the ISM is $\chi/\eta$, hence the overdamped limit corresponds to $\eta$ so large that the inertial scale is shorter than the microscopic time scale of the system (since the inertial $\chi$ is always an innocuous parameter of order $1$, the overdamped limit is often identified as that of large $\eta$). In this regime, the spin becomes a fast variable, so we can set $\dot{\boldsymbol s}_{i}=0$, and get, 
\begin{equation}
\frac{\eta}{\chi} \boldsymbol s_i = \boldsymbol v_i \times J \sum_j n_{ij} \boldsymbol v_j  + \boldsymbol \xi_i  \ ,
\end{equation}
which, once plugged into the equation for $\dot{\boldsymbol v}_{i}$, yields, 
\begin{equation}
\eta\frac{d \boldsymbol v_i}{dt}  =  \left(\boldsymbol v_i \times J \sum_j n_{ij} \boldsymbol v_j  + \boldsymbol \xi_i \right) \times \boldsymbol v_{i} \ ,
\end{equation}
which is exactly the Vicsek model \eqref{eq:vicsek}. From the conceptual point of view, this is particularly pleasing: the ISM is not some completely new model that has nothing to do with Vicsek; the ISM is simply the underdamped version of the Vicsek model. Or -- more accurately -- the Vicsek model is the overdamped limit of the ISM. Also, in this way we better understand where  the parameter $\eta$ in the Vicsek model comes from.

\section{The issue of dissipation and the RG crossover}
\label{sec:issue-dissipation-rg}

The route from an underdamped model for low $\eta$ to an overdamped one for large $\eta$ has a particularly clear and vivid representation in the context of the Renormalization Group (RG), where the two limits are described by two different asymptotic fixed points \cite{cavagna2023natural}. The crucial concept  is that in the vicinity of the underdamped fixed point ($\eta=0$), dissipation is a {\it relevant parameter} in the RG sense: this means that no matter how small $\eta$ is in the microscopic equations of motion, as long as it is non-vanishing, dissipation grows indefinitely along the RG flow, asymptotically reaching the overdamped limit ($\eta=\infty$). Because in the overdamped limit one recovers the Vicsek model, which has none of the new ISM features meant to describe natural flocks and swarms, one may wonder what is the purpose the ISM, and, most importantly, how is it that natural systems display clear underdamped features if the RG dictates that dissipation always takes over in the asymptotic limit?

The key word here is {\it asymptotic}. The fact that the limit of the RG flow is a certain fixed point means that the {\it infinite length scales} physics of that system is ruled by that fixed point; the process of iterating the RG transformation, i.e.\ to advance the RG flow, corresponds to observing the system at larger and larger length scales, so the attractive fixed point rules the full system in its thermodynamic limit.  However, biological groups are invariably far from the thermodynamic limit: they are {\it finite-size} systems, in which case it becomes essential to take into account the possibility of a RG crossover.

Consider a microscopic theory whose physical (i.e.\ bare) parameters are close to some RG fixed point $A$ and consider the case in which $A$ is {\it unstable} with respect of one of these parameters.  To anticipate what happens in our case, let us call $\eta$ this unstable parameter, and $\eta=0$ its value at the fixed point $A$. The RG flow along the direction $\eta$ leads from $A$ to some attractive fixed point $B$, with respect to which $\eta$ is a stable (irrelevant) direction.
Which fixed point, $A$ or $B$, will rule the physics of the system?  This depends on two things: the microscopic (physical) value of $\eta$ and the system's size $L$. The microscopic value of $\eta$ determines a {\it crossover length-scale}, ${\cal R}_c(\eta)$, below which $A$ still rules the physics of the system and beyond which $B$ does.  ${\cal R}_c(\eta)$ is the larger the smaller is $\eta$ and goes to infinity when $\eta=0$ (as it should, since killing kill the unstable
parameter makes $A$ an attractive fixed point).  Hence, if we are at liberty to take the thermodynamic limit, as some point we will necessarily cross-over to the regime $L \gg {\cal R}_c$ and the stable fixed point $B$ will therefore be dominant in that limit. On the contrary, if the system has finite size and it happens that $L \ll {\cal R}_c$ (and recall that ${\cal R}_c(\eta)$ can be arbitrarily large provided $\eta$ is small enough), the system is dominated by the unstable fixed point $A$. 

Several studies, both experimental and theoretical \cite{attanasi2014information, cavagna2017swarm, cavagna2023natural}, have shown that bird flocks and insect swarms have a small enough microscopic dissipation $\eta$ to be in the vicinity of the underdamped fixed point, namely $L \ll {\cal R}_c(\eta)$. Numerical simulations of the ISM have confirmed this scenario \cite{cavagna2019short, cavagna2019long}: 
if we run simulations of equations~\eqref{eq:ISM} at a given size $L$ with a small enough value of $\eta$, we find the critical exponents of the underdamped fixed point ($\eta=0$), while if we increase $\eta$ in the simulation at some point we cross over to the critical exponents of the overdamped fixed point, namely of the Vicsek model.  So, even though we introduced the ISM to reproduce some inertial features of the dynamics, it is all right to introduce relaxation through a pair of dissipative fluctuation-dissipation terms as in~\eqref{eq:ISM}, because for small enough $\eta$ we still recover the underdamped physics of real biological groups. However, there is more to say about the nature of the $\eta=0$ fixed point.

\subsection{The need of fully conservative fluctuation-dissipation}

What kind of fixed point is that with $\eta=0$?  It is underdamped, clearly, but how can the system relax there?  If $\eta=0$, we see from \eqref{eq:ISM} that the perfectly underdamped version of the ISM is the {\it deterministic} model~\eqref{eq:ISM-det}, which is somewhat disturbing, if not downright alarming. From the practical point of view this may seem not such a crucial issue, because --- as we have just explained --- we can always use some small-but-non-zero dissipation $\eta$ in the microscopic model and all will be good as long as $L \ll {\cal R}_c(\eta)$. But from a theoretical point of view, we cannot help asking how can we make statistical-mechanical sense of the model with $\eta=0$? The --- somewhat astounding --- answer is that the RG itself generates (under coarse-graining) a new pair of {\it conservative} fluctuation-dissipation terms, which makes the theory stochastic and relaxational even when $\eta=0$ \cite{cavagna2019short, cavagna2019long}. Let us see very briefly how this works. 

We start from a field-dynamical equation for the coarse-grained spin $\boldsymbol s(x,t)$ that is the continuous equivalent of the ISM equation, 
\begin{equation}
\frac{\partial \boldsymbol s(x,t)}{\partial t} = \mathrm{reversible \ terms} -\eta \boldsymbol s(x,t) + \boldsymbol \xi \ ,
\label{baccalu2}
\end{equation}
where the noise has correlator, 
\begin{equation}
\langle \xi_\alpha(x,t) \xi_\beta(x',t') \rangle = 2\eta \ \delta(x-x') \delta(t-t') \delta_{\alpha\beta} \ ,
\label{ap-pippazzu}
\end{equation}
and where the only important thing we need to know here about the reversible terms is that they are invariant under rotations of the velocity field and therefore conserve the total spin, just as the social force in the ISM; on the contrary, the two irreversible terms, friction and noise, break the conservation law. If the dynamical RG is applied to this theory, one finds that the irreversible terms get modified as follows \cite{cavagna2019short, cavagna2019long},
\begin{equation}
\frac{\partial \boldsymbol s(x,t)}{\partial t} = \mathrm{reversible \ terms} -(\eta-\lambda\nabla^2) \boldsymbol s(x,t) + \boldsymbol \xi \ ,
\label{ap-baccala}
\end{equation}
with,
\begin{equation}
\langle \xi_\alpha(x,t) \xi_\beta(x',t') \rangle = 2(\eta-\lambda\nabla^2) \ \delta(x-x') \delta(t-t') \delta_{\alpha\beta} \ ,
\label{ap-pippazza}
\end{equation}
This little piece of magic of the RG derives from the conservative nature (i.e. from the symplectic structure) of the reversible terms, which generate under coarse-graining an equally conservative fluctuation-dissipation structure \cite{cavagna2019short, cavagna2019long}.

Therefore, at the coarse-grained level (i.e.\ in the associated field theory) all is smooth and good at the underdamped, $\eta=0$, fixed point, because there is a conserved noise and a conserved `friction' that grant relaxation even in complete absence of dissipation. In fact, in the classical non-self-propelled (or fixed-network) case, this coarse-grained model is a very well-known field theory, namely Model G of Halperin-Hohenberg \cite{HH1977}, which is a perfectly conservative theory where no dissipation what-so-ever is present (that is, $\eta=0$ and $\lambda\neq 0$).  Bottom line: even though our microscopic model only sports dissipative fluctation-dissipation terms, its associated coarse-grained field theory generates non-dissipative fluctuation-dissipation terms that make it perfectly acceptable to talk about the underdamped fixed point at $\eta=0$. 

As pleasing (or mysterious) as this may seem, the fact remains that in the {\it microscopic} model, which is the only one we actually simulate numerically, setting $\eta=0$ remains problematic, because we do not have any $\lambda$ term to save the day. Whenever we need to test a theoretical result that holds at the underdamped RG fixed point corresponding to $\eta=0$, we {\it cannot} directly set $\eta=0$ in our simulation.  Instead, we rather have to calibrate $\eta$ and $L$ in such a way as to be able to relax the system while at the same time keeping $L \ll {\cal R}_c(\eta)$ in order to reproduce the underdamped phenomenology as observed e.g.\ in \cite{cavagna2023natural}.  This is realistically fine (as this is what happens in real natural system) but operationally very unpractical, not to say illogical: once we know that we need to test some feature of the underdamped fixed point, it would make more sense to just set $\eta=0$ also in the microscopic simulation.  How can we do that? The answer is: by carrying the conservative fluctuation-dissipation terms generated by the RG over to the microscopic theory.

\section{On the crossover length ${\cal R}_c$}
\label{sec:crossover-length}

We have stated that the system is underdamped when $L \ll {\cal R}_c$, where ${\cal R}_c$ is the $\eta$-dependent crossover length scale, but we did not give much insight on how this length scale emerges. Once we write \eqref{eq:cISM_ac+eta}, however, we can be a bit more specific.

The discrete Laplacian is dimensionless, so out of dimensional consistency we can write, $\Lambda_{ij} \sim - a^2\nabla^2$, where $a$ is some microscopic length-scale (it would be the lattice spacing, in the equilibrium case). 
From mere dimensional analysis, then, we see that the ratio $a^2\lambda/\eta$ has the dimensions of a square length; moreover, we see that such length-scale diverges for $\eta\to 0$, as we expect from the crossover length scale: if no dissipation is present, the system is underdamped over {\it all} scales. It is no surprise then to discover that {\it at the bare level}, i.e. without taking into account RG corrections, the crossover length-scale is indeed proportional to, 
\beq
{\cal R}^\mathrm{(bare)}_c \sim \sqrt{\frac{\lambda}{\eta}}  \ .
\label{dudelito}
\eeq
There are RG corrections to this formula, both in the amplitudes (to restore dimensions) and in the exponent \cite{cavagna2023natural}, but this simple argument gives flesh to the statement that finite systems are underdamped as long as dissipation is small enough.

\bibliography{COBBS-classic-bibliography_JAVIER}

\end{document}